\documentclass[12pt]{article}
\usepackage{xspace}
\usepackage{epsf}
\begin{document}
\title{Stochastic Resonance in a Periodically Modulated
Dissipative Nuclear Dynamics}
\author{ V.P.Berezovoj, Yu.L.Bolotin,
O.P.Dzyubak, V.V.Yanovsky, A.V.Zhiglo}
\maketitle
\begin{center} \small{\it{ Institute for Theoretical Physics of
National Science Center "Kharkov Institute\\ of Physics and
Technology" Akademicheskaya St. 1, Kharkov 61108, Ukraine}}
\end{center} \vspace{0.5cm}
\begin{abstract}
     A fission decay of highly excited periodically driven compound
        nuclei is considered in the framework of Langevin approach.
        We have used residual-time distribution (RTD) as the tool for studying of
         dynamic features in a presence of periodic perturbation. The
          structure of RTD essentially depends
          on the relation between Kramers decay rate and the frequency $\omega$ of
          the periodic perturbation. In particular, intensity of the first
          peak in RTD has a sharp maximum at certain nuclear temperature
          depending on $\omega$. This maximum should be considered as first-hand
          manifestation of stochastic resonance in nuclear dynamics.
\end{abstract}
\newcommand{\SR}{\mbox{SR}\xspace}\newcommand{\RTD}{\mbox{RTD}\xspace}
\newcommand{\MRT}{\mbox{MRT}\xspace}
\section{Introduction} The atomic nucleus since its discovery has been
constantly used for verifying of new physical ideas such as
tunneling \cite{1}, superfluids \cite{2}, superconductivity
\cite{3}, supersymmetry \cite{4}, dynamical chaos \cite{5}. Thus
it seems unnatural that one of most recent and intriguing
discoveries in nonlinear physics-stochasic resonance (\SR) (see
\cite{6} for a recent review) up till have not found response of
the nuclear community. This is particularly odd because there is
no doubt that the theory of the collective nuclear motion
pretending on a consistent description of nuclear dynamics must be
essentially nonlinear theory. The aim of the present work is to
demonstrate the principle possibility of observation of \SR in
nuclear dynamics. As a concrete example we consider a process of
induced nuclear fission in the presence of weak periodic
perturbation.

 \SR was introduced nearly 20 years ago to explain
the periodicity of the Earth's ice ages \cite{7,8} and has found
its numerous applications into such diverse fields like physics,
chemistry and biology (see \cite{6}).

 The mechanism of \SR can be explained in terms of
the motion of a particle in a symmetric double-well potential
subjected to noise and time periodic forcing. The noise causes
incoherent transitions between the two wells with a well-known
Kramers rate \cite{9} $r_k$. If we apply a weak periodic forcing
noise-induced hopping between the potential wells can become
synchronized with periodic signal. This statistical
synchronization takes place at the condition
\begin{equation}\label{1}
r_k^{-1} = \pi/\omega
\end{equation}
 where $\omega$ is a frequency of
periodic forcing. Two prominent feature of \SR arises from
synchronization condition (\ref{1}):

\qquad {\bf (i)}\  signal-to-noise ratio does not decrease with
increasing noise amplitude (as it happens in linear system), but
attains a maximum at a certain noise strength (optimal noise
amplitude can be found from (\ref{1}) as $r_k$ is simply connected
with it);

\qquad {\bf (ii)}\  the residence-time distribution (\RTD)
demonstrates a series of peaks, centered at odd multiples of the
half driving period $T_n = 2(n-\frac 1 2 )\frac{\pi}{\omega}$ with
exponentially decreasing amplitude. Notice that if a single escape
from a local potential well is the event of interest then \RTD
reveals the dynamics of considering system more transparently than
the signal-to-noise ratio. These signatures of \SR are not
confined to the special models,  but occur in general bi-~and
monostable systems and for different types of noise.

\section{Langevin Equation}

  Kramers \cite{9} was the first  to consider nuclear
fission as a process of overcoming the potential barrier by the
Brownian particle. A slow fission degree of freedom (with large
collective mass) is considered as Brownian particle, and fast
nucleon degrees of freedom --- as a heat bath. Adequacy of such
description is based on the assumption that the while of
equilibrium achievement in the system of nucleons degrees of
freedom is much less than the characteristic time scale of
collective motion. The most general way of description of
dissipative nuclear dynamics is Fokker-Planck equation \cite{10}.
However for demonstration of qualitative effects it is convenient
to use Langevin equation \cite{11} that is equivalent to
Fokker-Planck equation but is more transparent. As it has been
shown the description based on Langevin equation adequately
represents nuclear dissipative phenomena such as heavy-ion
reactions and fission decay \cite{12,13,14} and possesses a number
of advantages over Fokker-Planck description.

 Because we only intend to qualitatively
demonstrate \SR in nucleus let us consider the simplest type of
Langevin equation --- one-dimensional problem with inertial $M$
and friction $\gamma $ parameters independent on coordinates.
Fission coordinate $R$ is considered as a coordinate of Brownian
particle. The rest degrees of freedom play a role of heat bath
being modeled by random force $\xi(t)$.

The particle motion is described by Langevin equation for
canonically
 conjugate variables $\{ P,R\}$
\begin{eqnarray}\label{2}
\frac{dR}{dt} &=& \frac{P}{M} \nonumber \\ \frac{dP}{dt}&=&-\gamma
p-\frac{dV}{dR}+\xi(t) \\
 \beta&=&\gamma/M \nonumber
\end{eqnarray}

$\xi(t)$ is stochastic force possessing statistical properties of
white noise:
\begin{equation}\label{3}
\langle\xi(t)\rangle =0 ,\quad
\langle\xi(t)\xi(t')\rangle=2D\,\delta(t-t'), \quad D=\gamma T
\end{equation}
The nuclear temperature $T$(MeV)$=\sqrt{E^*/a}$ where $E^*$ is an
excitation  energy and the level density parameter $a=A/10$ ($A$
being a mass number). The (deformation) potential $V$ is given as
\cite{12}\\

\begin{equation}\label{4}
V(R)=\left\{ \begin{array}{ll} 37.46\,(R-1)^2\ (MeV) &{\mbox
{\qquad for }} 0<R<1.27\\
 8.0-18.73\,(R-1.8)^2\ (MeV) &{\mbox {\qquad for }} R>1.27 \end{array}\right.
\end{equation}
(these are parameters of $\: {}^{205}$At nucleus \cite{12}).

Plausible sources of periodic perturbation are considered below.

 The discretized form of the Langevin equation is
given by \cite{13,14}
\begin{equation}\label{5}
\begin{array}{rcl}
 R_{n+1} &=& R_n + \tau P_n/M \\
P_{n+1} &=& P_n(1-\beta \tau) - \left[\left( \frac{\textstyle
dV(R)}{\textstyle dR}\right)_n - A\cos\omega t_n\right]\tau +
\sqrt\frac{\textstyle 2\beta M
T\tau}{\textstyle N\mathstrut}\,\eta(t_n) %\label{5}
\end{array}\end{equation}
Here $t_n = n\tau$ and $\eta(t_n)$ is a normalized
Gaussian-distributed random variable which satisfies %<h(t)> = 0,

\begin{equation}\label{6}
\langle\eta(t)\rangle=0,\quad\langle\eta(t_n)\eta(t_{n'})\rangle=N\delta_{n
n'}
\end{equation}
Efficiency  of numerical algorithm  (\ref{5}) was checked  for the
following cases:

 \qquad{\bf (i)} \  $V = 0,\: A =0$, where  numerical and
analytical results for $\langle P^2\rangle$ and  $\langle
R^2\rangle$  can be compared \cite{12};

 \qquad {\bf (ii)} \
$V\not=0,\: A = 0$, where numerical and analytical values for
Kramers decay rate $r_k$ can be compared. According  to \cite{9}
\begin{equation}\label{7}
r_k = \frac{\omega_{min}}{2\pi}\left[ \sqrt{{\beta^*}^2 + 1} -
\beta^*\right] \,\exp(-{\Delta V}/T),\quad \beta^* =
\frac{\beta}{2\omega_{max}}
\end{equation}

Here $\omega_{min}$ and $\omega_{max}$ are the angular frequencies
of the potential (\ref{4}) at the potential minimum and
 at the top of barrier respectively, $\Delta V$ is the height of the
potential barrier. Numerical values of Kramers decay rates
${r_k}^i$ for the time bin $i$ is calculated by sampling the
number of fission events ${(N_f)}_i$ in the $i^{th}$ time bin
width $\Delta t$ normalized to the number of events $N_{total} -
\sum\limits_{j<i}{(N_f)}_j$ which have not fissioned
\begin{equation}\label{8}
 {r_k}^i = \frac{1}{N_{total} - \sum{(N_f)}_j}\frac{{(N_f)}_i}{\Delta t}
\end{equation}

Comparison of  (\ref{7}) with asymptotic value of (\ref{8}) was
used for determination of the time interval $\tau$, which provides
saturation for numerical integration (\ref{5}). On the other hand,
the interval $\tau$ should be chosen larger than the correlation
time of the random process $\xi(t)$. Results of numerical
calculations are plotted on Fig.1 according to (\ref{8}) under
different number of time steps per unit nuclear time
$\hbar/{MeV}$. One can see that  even 20 steps per nuclear time
provides a sufficient saturation.

%{\em Fig.1. The results of numerical calculations for  $r_k$ under
%different number of time steps per nuclear time  $\hbar/{MeV}$}.\\
%\begin{figure}\centering \framebox[15cm]{\rule[-15mm]{.1mm}{3cm}}
\begin{figure}[h]
\epsfxsize=13cm
\begin{center}
\epsffile{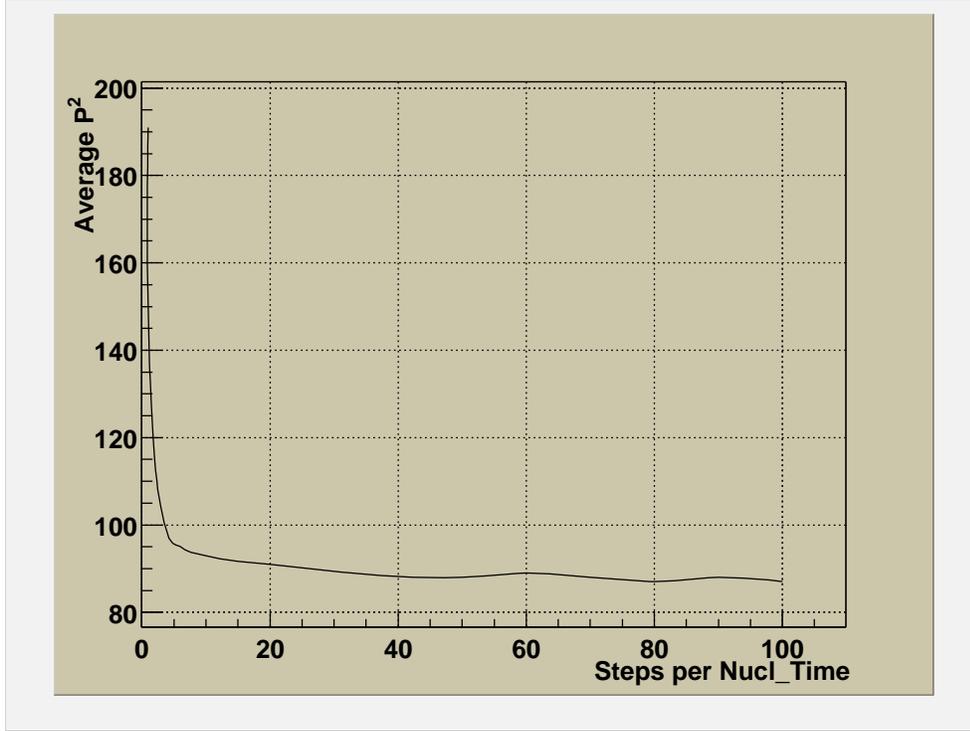}
%\caption{}
\end{center}
\caption{\protect \parbox[t]{12cm}{The results of numerical
calculations for $r_k$ under different number of time steps per
nuclear time $\hbar/{MeV}$}}\label{f1}\end{figure}

\section{Stochastic Resonance in Nuclear Fission}
 Now let us proceed to the description of expected effect
--- manifestation of \SR in nuclear fission. In the absence of
periodic forcing, \RTD $N(t)$ has the exponential form (see
\cite{6}) $N(t) ~ \exp(-r_k t)$. In the presence of the periodic
forcing, one observes a series of peaks, centered at odd multiples
of  the half driving period $T_\omega = 2 \pi/\omega$. The heights
of these peaks decrease exponentially with their order number.
These peaks are simply explained \cite{15}. The best time for the
particle to escape potential well is when the potential barrier
assumes a minimum. A phase of periodic perturbation may be chosen
in such a special way that the potential barrier $V(R)-A R
\cos(\omega t + \phi)$ assumes its first minimum at  $t =1/2\:
T_\omega$. Thus $t=1/2\: T_\omega$ is a preferred residence time
interval. Following "good opportunity" to escape occurs in a full
period, when potential barrier achieves its minimum again. The
second peak in the \RTD is therefore located at $3/2\: T_\omega$.
The location of the other peaks is evident. The peak heights decay
exponentially because the probabilities of the particle to jump
over a potential barrier are statistically independent. As is
shown  for symmetric double-well potential \cite{16}, the strength
$P_1$ of the first peak at $1/2\: T_\omega$ (the area under peak)
is a measure of the synchronization between the periodic forcing
and the switching between the wells. So, if the mean residence
time (\MRT) of the particle in one potential well is much larger
than the period of the driving, the particle is not likely to jump
over the first time the relevant potential barrier assumes its
minimum. The \RTD exhibits in such a case a larger number of peaks
where $P_1$ is small. If the \MRT is much shorter than the period
of the driving \RTD has already decayed practically to zero before
the time $1/2\: T_\omega$ is reached and the weight $P_1$ is again
small. Optimal synchronization, i.e., maximum $P_1$, is reached
when the \MRT matches half driving period, i.e., condition
(\ref{1}). This resonance condition can be achieved by varying the
noise intensity $D$ (or $\omega$).

 We will show that the same correlation between periodic forcing and
escape time takes place for a decay of excited states (fission)
with a single potential minimum as well. For \RTD constructing
(and following $P_1$ calculation) we use the numerical solutions
of Langevin equation (\ref{5}). Let us study evolution of  $P_1$
within the temperature interval $1\: MeV\leq T\leq 6\: MeV$.
Corresponding Kramers rates $r_k$ and resonance frequency
satisfying (\ref{1}) are represented in Table~1. Let us fix a
frequency of periodic perturbation $\omega = 0.0267 MeV/\hbar$
($T_\omega/2 = 117\hbar/MeV$) --- a resonance frequency at $T =
3MeV$ (see Table~1). On account of the exponential decay of peaks
heights in \RTD ($H_n \sim \exp (-r_k T_n),\; T_n =
2(n½1/2)\,\pi/\omega$) , one must observe a series of resonance
peaks at $T < 3 MeV$. On the other hand, at $T > 3 MeV$ (and for
the same frequency of periodic perturbation) vast majority of
nuclei would decay in a while shorter than $T_1 \sim T_\omega /2$.
Due to this a sharp maximum of first peak intensity should be
observed in the vicinity of $T \sim 3 MeV$, that is to be
interpreted as a manifestation of \SR.
 \begin{table}\centering \begin{tabular}{|c|c|c|}\hline
 \bf Temperature&\bf Kramers decay &\bf Resonant\\
 \bf ($MeV$)&\bf rate ($MeV/\hbar$)&\bf frequency($MeV/\hbar$)\\
 \hline 1 & $4.13\,\cdotp 10^{-5}$ & $1.3\,\cdotp 10^{-4}$\\
 2 & 0.0023 & 0.007\\ 3 & 0.0085 & 0.027\\4 & 0.0166 & 0.050\\
 5 & 0.0248 & 0.074\\ 6 & 0.0324 & 0.097\\ \hline
 \end{tabular}\caption{}\label{t1}\end{table}

The results of numerical procedure for \RTD are presented on
Fig.2. Pictures correspond to 
% to low ($T=1$ MeV), and (2b) --- to high ($T=6$ MeV) 
%
values of Kramers decay rate ( $T=$ 1 --- 6 MeV )
under fixed parameters of
periodic perturbation ($A = 1, \omega = 0.0267$). In accordance
with expected behavior in the first case (at low $r_k$~) one can
distinctly see three peaks located near $t = T/2 (\sim 117.7),
3/2T (\sim 353), 5/2T(\sim 588)$, and in the second case almost
all \RTD is concentrated near $ t = 0$ (with width less than
$T/2$). Connected with these variations of 1st peak intensity
(that represents the measure of the synchronization between the
periodic forcing and the nuclear temperature and consequently
measure of \SR) are depicted on Fig.3 for two frequencies of
periodic perturbation. Maxima of intensities $P_1$ coincide with
chosen frequencies of periodic perturbation (see \ref{t1}).

\begin{figure}[h]
\epsfxsize=13cm
\begin{center}
\epsffile{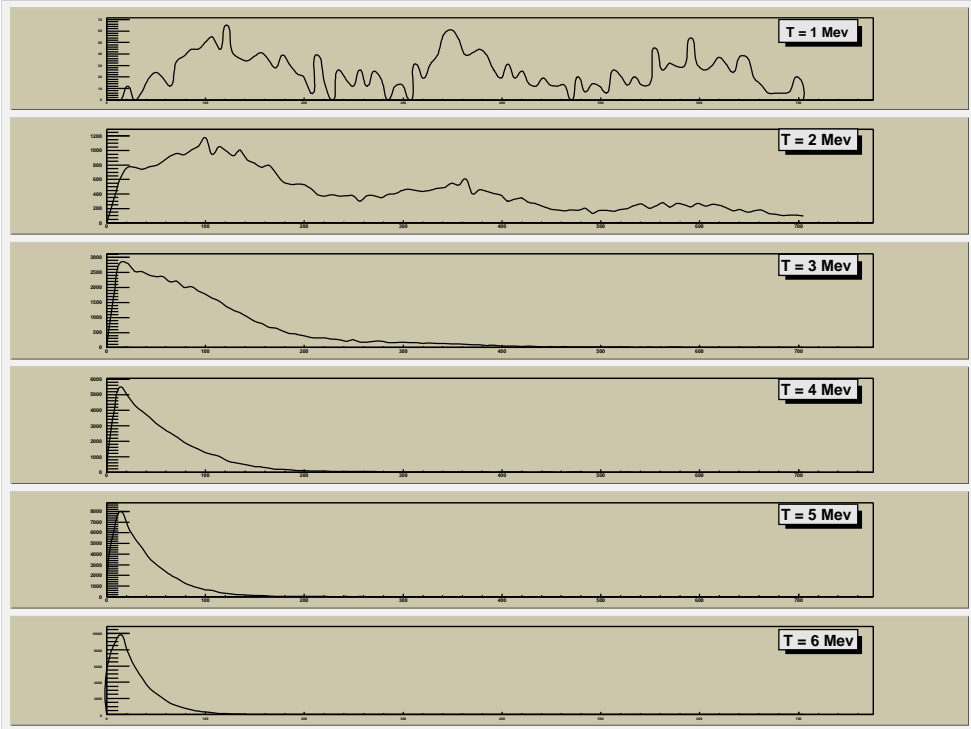}
%\caption{}
\end{center}
\caption{RTD for $T$ = 1 --- 6 $MeV$}
\label{f3}\end{figure}

%\begin{figure}\centering \framebox[15cm]{\rule[-15mm]{0mm}{2.8cm}}
%\caption{a) RTD for $T=1\: MeV$}\label{f2a}\end{figure}
%\addtocounter{figure}{-1}
%\begin{figure}\centering \framebox[15cm]{\rule[-15mm]{0mm}{2.8cm}}
%\caption{b) RTD for $T=6\: MeV$}\label{f2b}\end{figure}

\begin{figure}[h]
\epsfxsize=13cm
\begin{center}
\epsffile{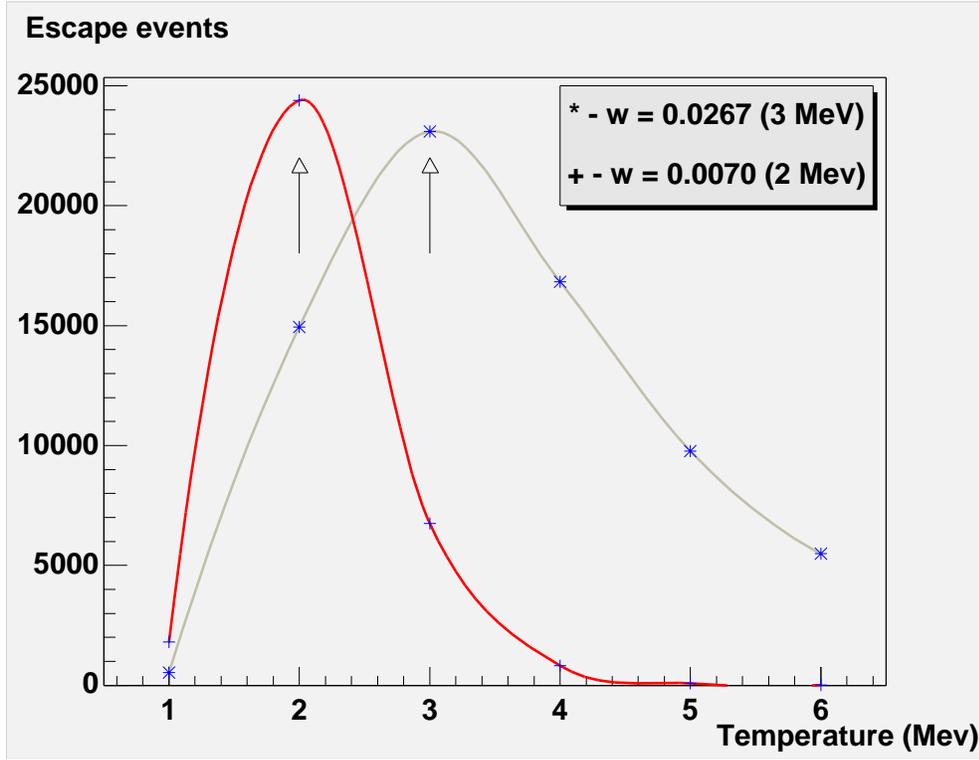}
%\caption{}
\end{center}
\caption{Dependence of $P_1$ on $T$ for two different $\omega$.
}\label{f3}\end{figure}

 In conclusion, let us briefly consider the possible sources  
of periodic perturbation. The first possibility is the fissile
nucleus as a component of double nuclear system formed, for
example, in heavy-ion collisions \cite{17}. In this case,
deformational potential will experience periodic perturbation
similar to tide-waves on the Earth caused by the Moon rotation. In
the case of asymmetric fission the source of periodic perturbation
may be alternating electric field. The problem of choice of
periodic perturbation would be discussed separately.

 We would like to thank A.Yu. Korchin for valuable discussions.

\end{document}